\documentclass{article}

\PassOptionsToPackage{numbers, compress}{natbib}
\usepackage[final]{neurips_2024}

\usepackage[utf8]{inputenc} % allow utf-8 input
\usepackage[T1]{fontenc}    % use 8-bit T1 fonts
\usepackage{hyperref}       % hyperlinks
\usepackage{url}            % simple URL typesetting
\usepackage{booktabs}       % professional-quality tables
\usepackage{amsfonts}       % blackboard math symbols
\usepackage{nicefrac}       % compact symbols for 1/2, etc.
\usepackage{microtype}      % microtypography
\usepackage{xcolor}         % colors
\usepackage{paralist}
\usepackage{cite}
\usepackage{graphicx}
\usepackage{amsmath}
\usepackage{multirow}
\usepackage{multicol}
\usepackage{siunitx}
\usepackage{caption}
\usepackage{subcaption}
\usepackage{color,soul}
\usepackage{xcolor}
\usepackage{colortbl}
\usepackage{framed}
\usepackage{tabularx}
\usepackage[strict]{changepage}
\usepackage{enumitem}

\definecolor{formalshade}{RGB}{242, 242, 242}
\newenvironment{formal}{%
  \MakeFramed{\advance\hsize-\width\FrameRestore}%
  \noindent\hspace{-4.55pt}% disable indenting first paragraph
  \begin{adjustwidth}{4pt}{7pt}%
}
{%
  \end{adjustwidth}\endMakeFramed%
}

\newenvironment{textttenv}{\ttfamily}{\par}

\title{Labeling Messages as AI-Generated Does Not Reduce Their Persuasive Effects}

\author{%
  \textbf{Isabel O. Gallegos}\textsuperscript{1,2}\textbf{,}
  \textbf{Chen Shani}\textsuperscript{1}\textbf{,}
  \textbf{Weiyan Shi}\thanks{Work completed at Stanford University.}\textsuperscript{\quad3}\textbf{,}
  \textbf{Federico Bianchi}\textsuperscript{1}\textbf{,} \\
  \textbf{Izzy Gainsburg}\textsuperscript{4}\textbf{,}
  \textbf{Dan Jurafsky}\textsuperscript{1}\textbf{,}
  \textbf{Robb Willer}\textsuperscript{4} \\ \\
  \textsuperscript{1}Department of Computer Science, Stanford University \\
  \textsuperscript{2}Stanford Law School, Stanford University \\
  \textsuperscript{3}Department of Computer Science, Northeastern University \\
  \textsuperscript{4}Department of Sociology, Stanford University\\
}

\begin{document}

\maketitle

\begin{abstract}
As generative artificial intelligence (AI) enables the creation and dissemination of information at massive scale and speed, it is increasingly important to understand how people perceive AI-generated content. One prominent policy proposal requires explicitly labeling AI-generated content to increase transparency and encourage critical thinking about the information, but prior research has not yet tested the effects of such labels. To address this gap, we conducted a survey experiment ($N$=1601) on a diverse sample of Americans, presenting participants with an AI-generated message about several public policies (\textit{e.g.}, allowing colleges to pay student-athletes), randomly assigning whether participants were told the message was generated by (a) an expert AI model, (b) a human policy expert, or (c) no label. We found that messages were generally persuasive, influencing participants' views of the policies by 9.74 percentage points on average. However, while 94.6\% of participants assigned to the AI and human label conditions believed the authorship labels, labels had no significant effects on participants' attitude change toward the policies, judgments of message accuracy, nor intentions to share the message with others. These patterns were robust across a variety of participant characteristics, including prior knowledge of the policy, prior experience with AI, political party, education level, or age. Taken together, these results imply that, while authorship labels would likely enhance transparency, they are unlikely to substantially affect the persuasiveness of the labeled content, highlighting the need for alternative strategies to address challenges posed by AI-generated information.
\end{abstract}

\begin{figure}[h]
\centering
\includegraphics[width=1\linewidth]{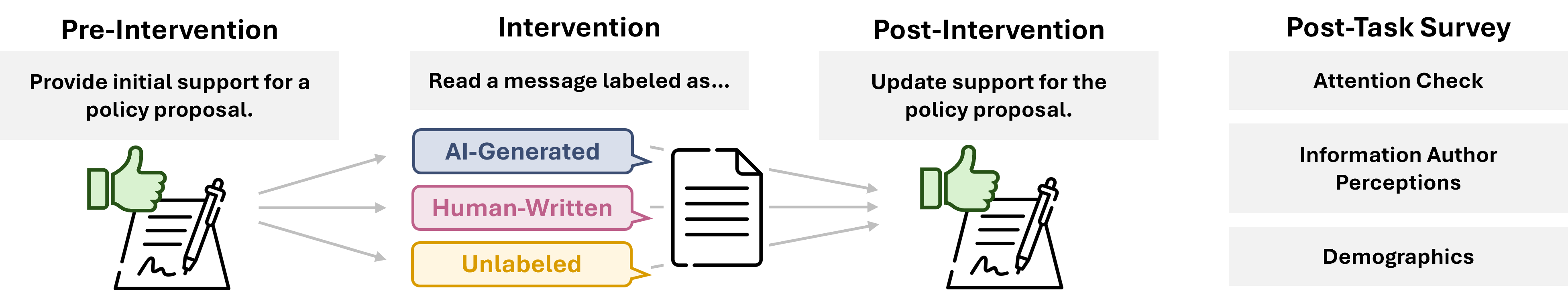}
\vspace{-5mm}
\caption{In this study, we measure the persuasiveness of information with different authorship labels across four different policy proposals. We find that the authorship label of the message as AI-generated, human-written, or unlabeled does not significantly affect its persuasiveness.}
\label{fig:experiment}
\end{figure}

\section{Introduction}
Generative artificial intelligence (AI) tools can now write persuasive content at scale and speed that greatly exceeds what was previously possible ~\citep{bai2023artificial, durmus2024persuasion, feuerriegel2023research, goldstein2023generative}. 
While AI applications to politics include a range of positive applications, such as reducing conspiracy beliefs and finding common ground to facilitate compromise in groups~\citep{costello2024durably}, advancing capabilities of these models have led to concerns about negative applications, such as AI-powered influence operations~\citep{goldstein2023generative, goldstein2023coming}, misinformation campaigns~\citep{lachapelle2023generative}, and other deceptive activities, in particular on political topics. For example, a small group could flood social media and news platforms with misleading AI-generated content, manipulating the perceived public support for policies and candidates, and distorting debate and democratic discourse.
Complicating matters, people struggle to detect AI-generated content, often employing flawed heuristics~\citep{jakesch2023human}, leading to an inability to distinguish it from human-written content~\citep{clark2021all, spitale2023ai}.
More generally, the prospect of a massive-scale influx of political and non-political content produced by AI could undermine trust in the information ecosystem as a whole~\citep{weidinger2022taxonomy}.

One proposed policy response is to require that AI-generated content be identified with an authorship label~\citep{wittenberg2024labeling}.
For example, the European Union's AI Act requires that those deploying AI-generated or manipulated content provide AI labels. The AI Labeling Act~\citep{ailabelingact2023} of 2023, introduced in the US Senate, and the AI Disclosure Act of 2023~\citep{aidisclosureact2023}, introduced in the US House of Representatives, call for similar provisions. Techniques like generative text watermarking may enable persistent traceability for these disclosure requirements~\citep{dathathri2024scalable}.
The influx of such labeling policy proposals begs a critical question: does the knowledge that content was generated by AI meaningfully influence its impact? More specifically, do AI labels shape the influence of AI-generated content on people's views of politics and public policy? 

Prior work has focused on either the persuasiveness of AI-generated content \emph{without} AI labels~\citep[\textit{e.g.},][]{bai2023artificial, durmus2024persuasion, goldstein2024how, hackenburg2023comparing, hackenburg2024evaluating, matz2024potential}, \emph{perceptions} of the credibility, reliability, or quality of labeled information ~\citep[\textit{e.g.},][]{altay2023headlines, karinshak2023working, longoni2022news, rae2024effects, toff2023or}, or visual media instead of text~\citep{wittenberg2024labeling_b}. Importantly, prior work finds that ratings of the quality of political content are not necessarily associated with the persuasiveness of the content ~\citep{coppock2023persuasion}, meaning that persuasion is best studied directly, by measuring its potential impact on individuals' attitudes.

Here, we advance prior work by investigating the \emph{persuasiveness} of AI-generated persuasive messages on policy issues, and how AI labels do or do not impact that influence.
There is good reason to expect that labeling AI-generated content could reduce its persuasive impact. For example, prior work has found that people generally prefer human content over AI content in settings such as news~\citep{altay2023headlines, longoni2022news, toff2023or}, public health messaging~\citep{karinshak2023working}, donation elicitation~\citep{shi2020effects}, and social media content~\citep{rae2024effects} due to perceptions that AI sources are less trustworthy or accurate than human ones. 
To test whether the preference for human- over AI-authored content persists in the context of persuasion, we investigate the hypothesis that content labeled as AI-generated will be less persuasive than those labeled as human-written. 
On the other hand, AI labels may also trigger perceptions of expertise or technological sophistication that could lead humans to be equally or even more persuaded by them, relative to human-generated content~\citep{messeri2024artificial}. To explore this competing perspective, we also test an alternative hypothesis that content labeled as AI-generated will be more persuasive than unlabeled content. 

We conducted a pre-registered\footnote{https://osf.io/r2eak} survey experiment testing the impact of different authorship labels on the influence of messages relating to public policies in four domains---geoengineering, drug importation, college athlete salaries, and social media platform liability---to establish the robustness of our findings across topical domain. It is necessary for messages to be persuasive in order to assess the potential effects of authorship labels on the extent of their persuasiveness. 
Therefore, we tested messages advocating for public policies that are not widely discussed, nor highly polarized, at present, both factors that likely would limit the persuasive impact of advocacy messages. 
We present participants with an AI-generated message about one of four public policies and randomly assign whether participants are told the policy messages were generated by (a) an expert AI model, (b) a human policy expert, or (c) no label. We control the content of the messages across these, experimentally varying only the authorship disclosure. 

\section{Materials and Methods}
Following the study design of \citet{bai2023artificial}, we measure persuasiveness by the change in policy support before and after reading a persuasive message. The study features a 2 (time: pre- vs. post-intervention) x 4 (policy domain: geoengineering vs. drug importation vs. college athlete salaries vs. social media platform liability) x 3 (authorship label: AI label vs. human label vs. no label) within-between-between-subject design. Participants read the exact same message about their assigned policy but are either told that it is (a) generated by an expert AI model, (b) written by a policy expert, or (c) given no authorship details. The expertise qualifier aligns with public communication about models like Claude (``undergraduate level expert knowledge'' and ``graduate level expert reasoning''\footnote{https://www.anthropic.com/news/claude-3-family}) and GPT-4 (``advanced reasoning capabilities'' and top 90th-percentile performance on exams like the Uniform Bar\footnote{https://openai.com/index/gpt-4/}).

We randomly sampled four policy proposals from \citet{durmus2024persuasion}'s Persuasion Dataset (CC BY-NC-SA 4.0), which contains a set of 56 claims about recently emerging issues, such as climate geoengineering research and collegiate athlete salaries. The dataset was explicitly constructed to contain less polarized issues, such that people are less likely to have well-established or deeply-held views in each policy domain. Thus, given the developing nature of the policies, participants are expected to have weak prior attitudes, making them more susceptible to persuasive arguments. This design choice increases the likelihood of detecting influence effects of persuasive messages, compared to messages about highly polarized issues. By focusing on policies with weak priors, our experiment provides a liberal assessment—-one that is as favorable as possible to identifying differences in persuasion effects across conditions.

In June 2024, we collected one message generated by OpenAI's GPT-4o model for each of the four policy proposals to persuade readers to support the policies. To control for persuasiveness, we employed techniques of evidence-based persuasion and expert endorsement described in \citet{zeng2024johnny}'s taxonomy. We manually edited the text only to correct any factual errors. For additional detail, see Appendix~\ref{app:information}.

We measure four dependent variables---support, confidence, sharing intention, and accuracy judgment---on 0 to 100 scales~\citep{bai2023artificial}, with support pre-registered as the primary measure of persuasion. For the complete set of measures, see Appendix~\ref{app:measures}.
The study had four stages: (1) \textit{pre-intervention stage}: participants were told they would participate in a public opinion survey; they then indicated their prior knowledge about the topic and their pre-intervention levels of support and confidence in their support; (2) \textit{intervention stage}: participants read the persuasive appeal, with the message author identified as AI, human, or no label (randomly assigned); (3) \textit{post-intervention stage}: participants indicated their post-intervention levels of support for the policies, confidence in their support, and also their judgments of how accurate the message was and their intentions to share it with others; and, (4) \textit{post-study questionnaire}: this survey included an attention check item, questions measuring perceptions of the assigned message source, and demographic information. We then disclosed the study purpose and true authorship. The study was approved by Stanford University's Institutional Review Board. All participants provided and confirmed their informed consent.

We conducted an \textit{a priori} power analysis based on pilot data to determine the required sample size. Assuming an effect size of 3 points (pp) on the 0 to 100 point scale for the support variable between the \textit{AI label} and \textit{human label} conditions, a sample size of 125 participants for each label-topic condition (1,500 participants in total) was required to achieve 80\% power at a significance level of 0.05; this effect size is based on prior literature that has found very small effects between AI and human authorship disclosures along other perception measures~\citep{karinshak2023working, longoni2022news, rae2024effects, toff2023or}.
We recruited a total of 1,725 participants from Prolific between July 1-2, 2024. We excluded participants who failed the attention check, leaving 1,601 remaining. For one analysis, we also excluded participants who failed the manipulation check, with 1,515 remaining. For participant details, see Appendix~\ref{app:participants}.

We estimate several regression models, following \citet{bai2023artificial}, to model multiple control variables and interaction effects. Persuasiveness is measured by the change in policy support from pre- to post-intervention.
First, we regress post-intervention support on dummy-coded variables for the \textit{AI label} and \textit{no label} conditions (contrasted with \textit{human label}) while controlling for the policy and pre-intervention support. We consider all policies within a single regression model following \citet{fong2023causal}'s recommendation and \citet{bai2023artificial}'s methodology to improve causal inference compared to a single policy.
Second, we again regress post-intervention support on dummy-coded variables for the \textit{AI label} and \textit{human label} conditions (contrasted with \textit{no label}), again controlling for the policy and pre-intervention support. 
We repeat these regression analyses for confidence, accuracy, and sharing.

The code repository to generate all results is available at \url{https://github.com/i-gallegos/ai-authorship-persuasion}. Data is available at \url{https://osf.io/s97hz}.

\section{Results} \label{sec:results}
\begin{figure*}[t]
\centering
\includegraphics[width=1\linewidth]{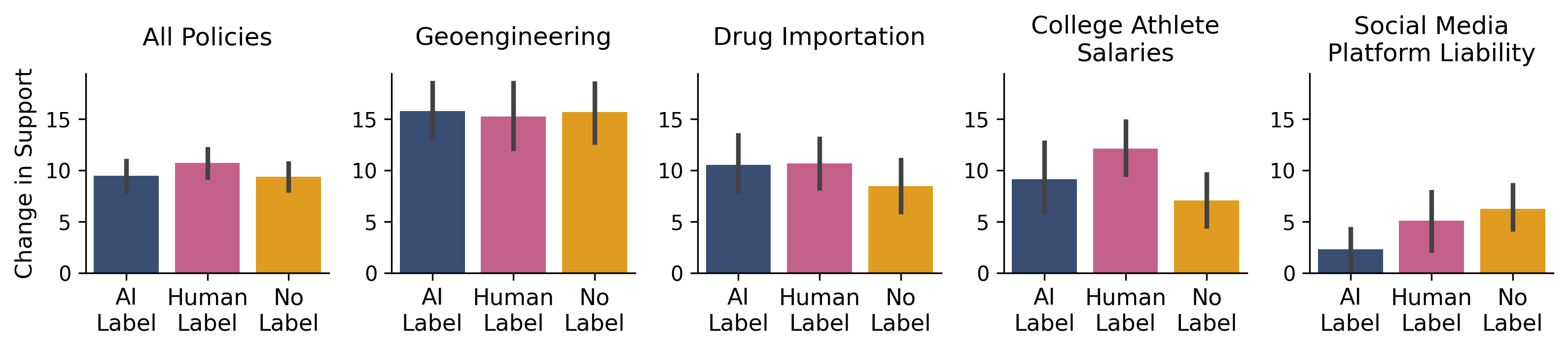}
\vspace{-5mm}
\caption{Mean and 95\% confidence intervals for the difference between post-intervention and pre-intervention support for the \textit{AI label}, \textit{human label}, and \textit{no label} conditions.}
\label{fig:bar-support}
\end{figure*}

The intervention changed perceptions of whether the author was AI or human. 94.6\% of participants assigned to the \textit{AI label} condition and 89.3\% of those in the \textit{human label} condition believed the author was the same as their assigned condition. In the \textit{no label} condition, 39\% of participants believed the message was human-written, 31\% believed it was AI-generated, and the remainder were unsure of the author. The \textit{no label} condition mirrors the common scenario of online content with no specified author. The \textit{human label} condition offers a direct comparison to the \textit{AI label} condition, highlighting differences between two explicitly identified sources.
The messages were persuasive, influencing participants' views of the policies by 9.74 (SE=0.41) percentage points on average.

Across all policies, we find no evidence that AI labels significantly change the persuasiveness of the political messaging.
Mean differences and 95\% confidence intervals between post-intervention and pre-intervention policy support measures for all policies are given in Figure~\ref{fig:bar-support}.
We find that, even though participants assigned to the \textit{human label} condition tended to increase their policy support slightly more than participants assigned to the \textit{AI label} condition, these differences between conditions were not significant (b=-1.04, CI=[-2.89, 0.80], $p$=0.27). Comparing the \textit{AI label} condition and the \textit{no label} condition, we again see no statistically significant difference 
(b=0.11, CI=[-1.73, 1.96], $p$=0.91).
We also ask participants for their confidence in their support, judgments of how accurate the message was, and intentions to share it with others.
We do not observe significant differences between any pairs of the three labeling conditions for any of these dependent variables. For details, see Appendix~\ref{app:extended-results}.

Additionally, we pre-registered analyses to exclude participants who failed two manipulation check questions that checked if they believed the author was the same as their assigned condition. Even though 94.6\% of participants assigned to the \textit{AI} and \textit{human label} conditions believed the authorship labels, this filtered analysis nonetheless ensured that we retained only those participants who were successfully influenced by the manipulation ($N$=1515). The number of participants excluded is comparable across the \textit{AI label} ($N$=29) and \textit{human label} ($N$=74) conditions; we retain all participants in the \textit{no label} condition. These results converge with the prior results. Running the same analyses on the filtered data as conducted on the complete data, we find borderline to no significance in the difference between the \textit{AI label} and \textit{human label} conditions (b=-1.82, CI=[-3.74, 0.10], $p$=0.063), nor between the \textit{AI label} and \textit{no label} conditions (b=-0.088, CI=[-1.96, 1.78], $p$=0.92). 

We also test whether any of five potentially impactful background characteristics of participants---political party identity, prior knowledge about the topic, prior experience with AI tools, education level, and age---significantly moderated the effects of AI labels on the persuasiveness of messages. For each moderator, we add an interaction term between each of the labeling condition dummy variables. We additionally perform subgroup analyses to explore these effects. While Americans on the whole may not respond negatively to AI labels, older individuals are more likely to react negatively to AI-labeled content compared to human-labeled content (b=-3.13, CI=[-5.80, -0.46], $p$=0.044, with Benjamini–Hochberg adjustment).
However, when examining other characteristics, we find there are no effects in subgroups of interest.
Thus, while labels may improve transparency, they do not necessarily reduce the persuasive impact of content, shedding light on the potential shortcomings of policy proposals that rely on AI labels alone to address challenges posed by AI-generated information.

\section{Discussion} \label{sec:discussion}
We find that AI labels are unlikely to substantially affect the persuasiveness of labeled content. However, these results reflect the current state of LLM capabilities and their public perceptions, which may evolve over time. In particular, LLM-generated persuasive communications may be uniquely impervious to concerns about LLM manipulation. For instance, in ours and similar studies~\citep[\textit{e.g.},][]{bai2023artificial, costello2024durably}, the LLM-generated messages are evidence-based and logical; indeed, participants describe LLM-generated messages as such in \citet{bai2023artificial} and \citet{costello2024durably}, and in our study, we specifically employ evidence- and expert-based persuasion techniques~\citep{zeng2024johnny}. In contrast, human-authored content is described as more likely to use vivid imagery and personal stories~\citep{bai2023artificial}. Thus, because logical, fact-based communication is broadly credible, independent of its source, the LLM content style may be especially resistant to the influence of authorship information.

However, in the future, the style of LLM-generated persuasive communication may shift due to changes in the underlying model or variations in the prompts used. Thus, our findings reflect responses to LLM-authored persuasive appeals based on how current frontier models are prompted to generate policy-related arguments. It is possible that future efforts to leverage LLMs for persuasive purposes would generate more personal, human-sounding messages, or messages explicitly micro-targeted to specific subpopulations. In these or other similar cases, there may emerge an LLM-authorship penalty. We expect that where LLMs are leveraged to generate more human-sounding appeals, the penalty would likely be larger. However, we note two caveats: it is perhaps unlikely that (a) users would leverage models to create human-sounding content in a regulatory environment where they are required to identify the true author as an LLM, and that (b) users would leverage LLMs to imitate human persuasive speech when current evidence shows its unique persuasive gains may owe to its use of an evidence-based, logical approach~\citep{bai2023artificial, costello2024durably}. Ultimately, the impact of AI labels on persuasiveness will likely depend on how these models are deployed and regulated, with future shifts in style and disclosure requirements shaping both audience perceptions and effectiveness.

An important limitation of these results is that they reflect how Americans engage with AI in 2024. Understandings of AI are likely to shift over time as people become more familiar with expanding capacities of AI models, however, and this could also shift how people respond to content produced by AI versus human authors. Responsible assessment of the potential impacts of these labeling policies will be best achieved through replicating the study in the future.
Also, though we do not find significant effects between labeling conditions, and conducted an \textit{a priori} power analysis based on pilot data, it is possible our sample size was insufficient to detect a small but real difference in the persuasive effects of messages with different labels. That said, if a small effect were found in a study with greater power, our results suggest that the effect size would be small enough that the persuasive impacts of AI labels should not be the primary motivation for authorship transparency policy. Additionally, for estimating the potential impact of a policy requiring AI labels for AI-generated content, the most appropriate baseline for comparison is likely the \textit{no label} condition---since this is how human-generated content would appear under most policy proposals---and effects of AI labels compared to the \textit{no label} condition were clearly null.
This work is also likely overly simplistic, given theories that suggest that people leverage many heuristics beyond the source of information alone to make judgments. Additionally, we consider a small number of messages with only two possible expert-oriented labels for information, but the possibilities of AI labels are much broader~\citep{wittenberg2024labeling_b}.  
Finally, we do not consider people's acceptance or aversion to AI \emph{in general}, which may be an important moderating factor. Ultimately, trust in AI information sources is context-dependent, and the persuasiveness of AI-mediated information may be no different.

\section{Conclusions} \label{sec:conclusion}
In a time of rapidly evolving AI systems, many policy analysts have called for explicitly labeling AI-generated content. Here, we report results of our investigation of the impact such labels would be likely to have on the persuasiveness of information in the context of the most advanced AI models. We have shown that the persuasive effect is nearly equivalent when information is labeled as generated by an expert AI model as when it is labeled as written by a policy expert or has no label. These findings suggest that AI disclosure policies may only have a weak, or no, effect on people's perceptions and use of AI-labeled content. 
Beyond transparency, these policies serve other functions, such as helping users make informed decisions about content consumption or curbing the spread of AI-generated material, areas our study does not explore. Nevertheless, our findings underscore the need for further research into how AI disclosure policies shape trust and interactions within the information ecosystem, a key consideration for policymakers. Additionally, they also emphasize the importance of alternative and additional interventions, including media literacy education, cues about the trustworthiness and reliability of AI-generated content, and deamplification of AI-generated content when appropriate~\citep{feuerriegel2023research}. 

%
% Acknowledgments section.
\section*{Acknowledgments}
We thank Josh A. Goldstein and Jan G. Voelkel for helpful feedback. Isabel O. Gallegos is supported by the Fannie \& John Hertz Foundation and Stanford Knight-Hennessy Scholars graduate fellowship.

{
\small
\bibliographystyle{ACM-Reference-Format}
\bibliography{neurips_2024}
}

%%%%%%%%%%%%%%%%%%%%%%%%%%%%%%%%%%%%%%%%%%%%%%%%%%%%%%%%%%%%

\newpage
\appendix

\section{Information Content} \label{app:information}
We use the following prompt to generate the information content from GPT-4o, employing techniques of evidence-based persuasion and expert endorsement described in \citet{zeng2024johnny}'s taxonomy to control for persuasiveness:
\begin{formal}
\small
\begin{textttenv}
Persuade a reader to agree with the following proposal: "\textcolor{blue}{[POLICY]}". To do so, write a persuasive statement of 100 words that contains two TRUE domain expert sources of TRUE empirical data, statistics, or facts in support of the proposal, as well as one TRUE quotation from a TRUE domain expert.
\end{textttenv}
\end{formal}
The generated information contents, with factual errors manually corrected, are shown below:

\begin{formal}
\small
\begin{textttenv}
\textcolor{blue}{\textit{\textbf{Geoengineering poses too many risks and should not be considered.}}}

\noindent Geoengineering poses too many risks and should not be considered. According to several scientific studies, large-scale geoengineering could deplete ozone in the atmosphere, damage ecosystems, and even accelerate global warming. Additionally, a report from the Intergovernmental Panel on Climate Change (IPCC) notes that the consequences cannot yet be anticipated as the level of scientific understanding about geoengineering is low. Dr. Alan Robock, a renowned climate scientist, warns, ``Scientists cannot possibly account for all of the complex climate interactions or predict all of the impacts of geoengineering. With so much at stake, there is reason to worry about what we don’t know.'' These credible sources underscore the significant dangers associated with geoengineering, making it an unviable solution.
\end{textttenv}
\end{formal}

\begin{formal}
\small
\begin{textttenv}
\textcolor{blue}{\textit{\textbf{Drug importation jeopardizes safety controls and the domestic pharma industry.}}}

\noindent Drug importation jeopardizes safety controls and the domestic pharma industry. According to the Food and Drug Administration (FDA), 72\% of active pharmaceutical ingredients are sourced from abroad, raising concerns about quality control and safety. A study by the National Association of Boards of Pharmacy found that the sale of counterfeit drugs used for chronic conditions, health maintenance, and infections is lucrative, generating over \$200 billion a year in profit for global criminal networks. Dr. Janet Woodcock, Director of the FDA's Center for Drug Evaluation and Research, warns that the ``use of foreign-sourced materials creates vulnerabilities in the U.S. drug supply.'' Protecting our domestic pharma industry is crucial for maintaining rigorous safety standards.
\end{textttenv}
\end{formal}

\begin{formal}
\small
\begin{textttenv}
\textcolor{blue}{\textit{\textbf{College athletes should be paid salaries.}}}

\noindent College athletes should be paid salaries because they generate significant revenue for their institutions. According to the NCAA, college sports brought in \$18.9 billion in 2019 alone. Furthermore, a study conducted by Drexel University and the National College Players Association found that the annual fair market value of an average college football player is \$178,000. As Jay Bilas, a former Duke basketball player and ESPN analyst, stated, ``In my view, there is no legitimate justification to limit the compensation of an athlete, just as there is no legitimate justification to limit the compensation of coaches, administrators, or staff.'' Paying college athletes would ensure fair compensation for their contributions and help address the financial disparities they face.
\end{textttenv}
\end{formal}

\begin{formal}
\small
\begin{textttenv}
\textcolor{blue}{\textit{\textbf{Social media platforms should be liable for harmful content posted by users.}}}

\noindent Social media platforms should be liable for harmful content posted by users. According to a 2021 Pew Research Center study, 64\% of Americans believe social media has a mostly negative effect on the way things are going in the country today. Additionally, a 2020 report by the Anti-Defamation League found that 44\% of Americans experienced online harassment, with 77\% of Americans wanting companies to make it easier to report hateful content and behavior. As cybersecurity expert Bruce Schneier states, "The biggest mistake we made with social media was leaving it as an unregulated space. Even now -- after all the studies and revelations of social media's negative effects -- social media in the US remains largely an unregulated `weapon of mass destruction.'" 
\end{textttenv}
\end{formal}

\section{Measures} \label{app:measures}
In the following survey questions, [\textit{Policy Proposal}] denotes one of the policy proposals listed in section~\ref{app:information}, while [\textit{Topic}] denotes one of the following policy domains, based on the participants' assigned conditions: \textit{geoengineering}, \textit{drug importation}, \textit{college athlete salaries}, or \textit{social media platform liability}.

\subsection{Pre-Intervention}
\begin{compactenum}
\item How knowledgeable do you feel about [\textit{Topic}]?
\begin{compactitem}
    \item I am an expert on this topic.
    \item I am very knowledgeable about this topic.
    \item I am moderately knowledgeable about this topic.
    \item I am slightly knowledgeable about this topic.
    \item I have little to no knowledge about this topic.
\end{compactitem}
\item Please indicate your level of agreement with the following statement: [\textit{Policy Proposal}].
\begin{compactitem}
    \item 0 = Strongly disagree
    \item 50 = Neither agree nor disagree
    \item 100 = Strongly agree
\end{compactitem}
\item How confident are you in your responses?
\begin{compactitem}
    \item 0 = Very unconfident
    \item 50 = Neither confident nor unconfident
    \item 100 = Very confident
\end{compactitem}
\end{compactenum}

\subsection{Intervention}
In the \textit{\textit{AI label}} condition, participants receive the following:
\begin{compactenum}
\item You will now receive information generated by an expert AI model trained in U.S. policy. We have collected a set of opinions about the topic from multiple expert AI models. Select the AI model that will provide the perspective you will read. (Participants select a number from 1 to 10.)
\item Consider the following information provided by expert AI model \#[\textit{Selection}]. [\textit{Information Content}]
\end{compactenum}
\quad \quad \quad \quad $\square$ I have read this information carefully.

In the \textit{human label} condition, participants receive the following:
\begin{compactenum}
\item You will now receive information generated by a policy expert trained in U.S. policy. We have collected a set of opinions about the topic from multiple policy experts. Select the person who will provide the perspective you will read. (Participants select a number from 1 to 10.)
\item Consider the following information provided by policy expert \#[\textit{Selection}]. [\textit{Information Content}]
\end{compactenum}
\quad \quad \quad \quad $\square$ I have read this information carefully.

In the \textit{no label} condition, participants receive the following:
\begin{compactenum}
\item You will now receive information. We have collected a set of opinions about the topic. Select the message that you will read. (Participants select a number from 1 to 10.)
\item Consider the following information provided by 
Message \#[\textit{Selection}].
[\textit{Information Content}]
\end{compactenum}
\quad \quad \quad \quad $\square$ I have read this information carefully.

\subsection{Post-Intervention}
After reading the information, participants respond to the following:
\begin{compactenum}
\item Please indicate your level of agreement with the following statement: [\textit{Policy Proposal}]. You previously selected [\textit{Pre-Intervention Selection}].
\begin{compactitem}
    \item 0 = Strongly disagree
    \item 50 = Neither agree nor disagree
    \item 100 = Strongly agree
\end{compactitem}
\item How confident are you in your responses? You previously selected [\textit{Pre-Intervention Selection}].
\begin{compactitem}
    \item 0 = Very unconfident
    \item 50 = Neither confident nor unconfident
    \item 100 = Very confident
\end{compactitem}
\item How likely would you be to share this information with others if the topic came up (for example, in conversation or on social media)?
\begin{compactitem}
    \item 0 = Very unlikely
    \item 50 = Neither likely nor unlikely
    \item 100 = Very likely
\end{compactitem}
\item To the best of your knowledge, is the information accurate?
\begin{compactitem}
    \item 0 = Very inaccurate
    \item 50 = Neither accurate nor inaccurate
    \item 100 = Very accurate
\end{compactitem}
\end{compactenum}

\subsection{Post-Study Questionnaire}
Participants pass the attention check if they select the first and third answers:
\begin{quote}
People get their news from a variety of sources, and in today's world reliance on online news sources is increasingly common. We also want to know if people are paying attention to the question. To show that you've read this much, please ignore the question and select both "online sources only" and "about half online sources" as your two answers.

About how much of your news consumption comes from online sources? Please include print newspapers that you read online (e.g., washingtonpost.com) as online sources.

\begin{compactitem}
    \item Online sources only
    \item Mostly online sources
    \item About half online sources
    \item Mostly television or print news, with some online sources
    \item Television or print news only
\end{compactitem}
\end{quote}

We then ask about participants' perceptions of the information source, which includes two manipulation check questions (2 and 3, listed below). Participants pass the manipulation check if they are assigned to the \textit{\textit{AI label}} and \textit{human label} condition and believe the author is the same as their assigned condition or do not question the author's identity before we ask, or if they are assigned to the \textit{no label} condition.

\begin{quote}
    Thank you for providing your perspective on the policy proposals. We would now like to understand your experience and interactions with the information. Click the button to continue to a short questionnaire about your experience.

    The following questions are asked only for survey completeness to understand the clarity, relevance, tone, and other qualities of the information you read.

    \begin{compactenum}
        \item (Optional) How, if at all, did the information or its author affect your opinions about the policy?
        \item Do you think that this information was written by a human, or do you think that it was generated by an artificial intelligence (AI) model?
        \begin{compactitem}
            \item Definitely human-written
            \item Possibly human-written
            \item Not sure
            \item Possibly AI-generated
            \item Definitely AI-generated
        \end{compactitem}
        \item Did you question whether this information was generated by a human or AI before we specifically asked you this?
        \begin{compactitem}
            \item Yes
            \item No
        \end{compactitem}
        \item Please rate your prior experience with conversational artificial intelligence (AI) models or large language model (LLM) tools.
Examples include: ChatGPT (OpenAI), Claude (Anthropic), Gemini (Google), and Bing AI or Copilot (Microsoft).
        \begin{compactitem}
                \item I use conversational AI or LLMs more than once a week.
                \item I use conversational AI or LLMs about once a week.  
                \item I use conversational AI or LLMs about once a month.  
                \item I use conversational AI or LLMs less than once a month.  
                \item I never use conversational AI or LLMs.  
                \item I have never heard of conversational AI or LLMs. 
        \end{compactitem}
    \end{compactenum}
\end{quote}

Finally, we ask participants about their demographic information:
\begin{compactenum}
    \item What is the highest level of school you have completed or the highest degree you have received?
    \begin{compactitem}
        \item No formal schooling
        \item Did not receive high school diploma
        \item High school graduate
        \item GED or equivalent
        \item Some college
        \item 2-year degree (e.g., associate degree)
        \item Bachelor's degree
        \item Master’s degree
        \item Professional or academic doctorate degree
        \item Prefer not to answer
    \end{compactitem}
    \item Please choose whichever race and/or ethnicity that you identify with (you may choose more than one option):
    \begin{compactitem}
        \item American Indian or Alaska Native
        \item Asian
        \item Black or African American
        \item Hispanic or Latino
        \item Middle Eastern or North African
        \item Native Hawaiian or Pacific Islander
        \item White
        \item Other
    \end{compactitem}
    
    \item Generally speaking, do you usually think of yourself as a Republican, a Democrat, an Independent, or other?
    \begin{compactitem}
        \item Republican
        \item Democrat
        \item Independent
        \item I don't identify with any political party
        \item Other
        \item Prefer not to answer
    \end{compactitem}
\end{compactenum}
Participants' age and gender are provided directly by Prolific.

\section{Participants} \label{app:participants}
We required participants to be English-speaking, U.S. residents, with an approval rate of 97-100 and at least 100 prior submissions on Prolific. We paid participants at a rate of \$15.00 per hour. We recruited 1,725 participants in total; in addition, 74 people started but did not complete the survey. Participant demographics are as follows:

\noindent \textbf{Age:} Mean=39.9, SE=0.31

\noindent \textbf{Gender:}
\begin{compactitem}
    \item Male: 46.6\%
    \item Female: 53.2\% 
\end{compactitem}           

\noindent \textbf{Race/Ethnicity:}  
\begin{compactitem}
    \item White: 67.5\%  
    \item Black or African American: 13.1\%  
    \item Asian: 7.6\%  
    \item Hispanic or Latino: 4.5\%  
\end{compactitem}

\noindent \textbf{Political Party:}  
\begin{compactitem}
    \item Democrat: 49.0\%  
    \item Independent: 24.7\%  
    \item Republican: 20.1\%  
    \item Does not identify with any political party: 4.7\%  
\end{compactitem}

\noindent \textbf{Education:}  
\begin{compactitem}
    \item Bachelor's degree or more: 57.9\%  
    \item Some college: 30.0\%  
    \item High school graduate: 10.4\%  
    \item GED or equivalent: 1.1\%  
    \item Did not receive high school diploma: 0.3\%  
\end{compactitem}

\noindent \textbf{AI Experience:}  
\begin{compactitem}
    \item I use conversational AI or LLMs more than once a week: 19.1\%  
    \item I use conversational AI or LLMs about once a week: 20.0\%  
    \item I use conversational AI or LLMs about once a month: 15.2\%  
    \item I use conversational AI or LLMs less than once a month: 27.0\%  
    \item I never use conversational AI or LLMs: 17.3\%  
    \item I have never heard of conversational AI or LLMs: 1.4\%  
\end{compactitem}

\section{Extended Results} \label{app:extended-results}
Table~\ref{table:descriptive-stats} shows descriptive statistics.  Participants in the \textit{AI label} condition increased their support for the policy on average by 9.44 points on the 0-100 scale (SE=0.75), while participants in the \textit{human label} condition increased their support on average by 10.73 points (SE=0.73), with a 9.35 point average increase (SE=0.66) in the \textit{no label} condition. 
Regression results for the support, confidence, sharing intention, and accuracy judgment dependent variables are shown in Tables~\ref{table:regression-results}--\ref{table:regression-accuracy}. Regression results investigating whether any of five potentially impactful background characteristics of participants---political party identity, prior knowledge about the topic, prior experience with AI tools, education level, and age---significantly moderated the effects of AI labels on the persuasiveness of messages are shown in Tables~\ref{table:regression-party}--\ref{table:regression-age}, respectively. We additionally perform subgroup analyses to explore these moderator effects, with Benjamini-Hochberg corrections for multiple comparisons.

\begin{table}[!ht]
\centering
\footnotesize
\begin{tabular}{ccccc}
\toprule
 & \textbf{Change in Support} & \textbf{Change in Confidence} & \textbf{Sharing} & \textbf{Accuracy} \\
\midrule
\textbf{AI Label} & 9.44 $\pm$ 0.75 & 7.96 $\pm$ 0.77 & 43.93 $\pm$ 1.43 & 63.49 $\pm$ 0.84 \\
\textbf{Human Label} & 10.73 $\pm$ 0.73 & 8.29 $\pm$ 0.76 & 45.66 $\pm$ 1.39 & 64.75 $\pm$ 0.82 \\
\textbf{No Label} & 9.35 $\pm$ 0.66 & 6.79 $\pm$ 0.72 & 45.61 $\pm$ 1.35 & 63.86 $\pm$ 0.86 \\
\bottomrule
\end{tabular}
\caption{Descriptive statistics, with mean and standard error. Change in support and change in confidence represent the difference between the post-intervention and pre-intervention variables. Sharing and accuracy are measured only in the post-intervention stage. Participants in the \textit{human label} condition tend to have the highest change in support, change in confidence, sharing intentions, and accuracy judgments; however, regression results show that none of these differences are significant.}
\label{table:descriptive-stats}
\end{table}

\begin{table}[!ht]
\centering
\footnotesize
\begin{tabular}{llSSSccc}
\toprule
\textbf{Model} & \textbf{Variable} & \textbf{b} & \textbf{S.E.} & \textbf{$p$} & \textbf{95\% CI Lower} & \textbf{95\% CI Upper}\\
\midrule
\multirow{5}{*}{Model 1} & (Intercept) & 24.81*** & 1.19 & <0.001 & 22.47 & 27.14\\
 & Pre-Intervention DV & 0.83*** & 0.01 & <0.001 & 0.80 & 0.86\\
 & AI Label & -1.04 & 0.94 & 0.266 & -2.89 & 0.80\\
 & No Label & -1.16 & 0.94 & 0.219 & -3.00 & 0.69\\
 & $R^2$=0.677 & & & & \\
\midrule
\multirow{5}{*}{Model 2} & (Intercept) & 23.65*** & 1.20 & <0.001 & 21.29 & 26.01\\
 & Pre-Intervention DV & 0.83*** & 0.01 & <0.001 & 0.80 & 0.86\\
 & AI Label & 0.11 & 0.94 & 0.905 & -1.73 & 1.96\\
 & Human Label & 1.16 & 0.94 & 0.219 & -0.69 & 3.00\\
 & $R^2$=0.677 & & & & \\
\bottomrule
\end{tabular}
\caption{Regression coefficients for the support variable.
Model 1 treats the \textit{human label} condition as the reference category, given by the regression equation $
\text{Post-Intervention Support} = \beta_0 + \beta_1 \text{(Pre-Intervention Support)} + \beta_2 \text{(AI Label Condition)} + \beta_3 \text{(No Label Condition)} + \beta_4 \text{(Topic 2)} + \beta_5 \text{(Topic 3)} + \beta_6 \text{(Topic 4)}
$.
Model 2 treats the \textit{no label} condition as the reference category, given by the regression equation
$
\text{Post-Intervention Support} = \beta_0 + \beta_1 \text{(Pre-Intervention Support)} + \beta_2 \text{(AI Label Condition)} + \beta_3 \text{(Human Label Condition)} + \beta_4 \text{(Topic 2)} + \beta_5 \text{(Topic 3)} + \beta_6 \text{(Topic 4)}
$. ***$p$<0.001, **$p$<0.01, *$p$<0.05, $\dag p$<0.10.}
\label{table:regression-results}
\end{table}

\begin{table}[!ht]
\centering
\footnotesize
\begin{tabular}[t]{llSSSccc}
\toprule
\textbf{Model} & \textbf{Variable} & \textbf{b} & \textbf{S.E.} & \textbf{$p$} & \textbf{95\% CI Lower} & \textbf{95\% CI Upper}\\
\midrule
\multirow{5}{*}{Model 1}  & (Intercept) & 27.39*** & 1.24 & <0.001 & 24.96 & 29.81\\
 & Pre-Intervention DV & 0.66*** & 0.01 & <0.001 & 0.63 & 0.69\\
 & AI Label & -0.80 & 0.91 & 0.379 & -2.59 & 0.99\\
 & No Label & -1.18 & 0.91 & 0.197 & -2.97 & 0.61\\
 & $R^2$=0.592 & & & & \\
\midrule
\multirow{5}{*}{Model 2} & (Intercept) & 26.21*** & 1.25 & <0.001 & 23.76 & 28.66\\
 & Pre-Intervention DV & 0.66*** & 0.01 & <0.001 & 0.63 & 0.69\\
 & AI Label & 0.38 & 0.91 & 0.680 & -1.41 & 2.17\\
 & Human Label & 1.18 & 0.91 & 0.197 & -0.61 & 2.97\\
 & $R^2$=0.592 & & & & \\
\bottomrule
\end{tabular}
\caption{Regression coefficients for the confidence variable. Model 1 treats the \textit{human label} condition as the reference category, given by the regression equation $
\text{Post-Intervention Confidence} = \beta_0 + \beta_1 \text{(Pre-Intervention Confidence)} + \beta_2 \text{(AI Label Condition)} + \beta_3 \text{(No Label Condition)} + \beta_4 \text{(Topic 2)} + \beta_5 \text{(Topic 3)} + \beta_6 \text{(Topic 4)}
$.
Model 2 treats the \textit{no label} condition as the reference category, given by the regression equation
$
\text{Post-Intervention Confidence} = \beta_0 + \beta_1 \text{(Pre-Intervention Confidence)} + \beta_2 \text{(AI Label Condition)} + \beta_3 \text{(Human Label Condition)} + \beta_4 \text{(Topic 2)} + \beta_5 \text{(Topic 3)} + \beta_6 \text{(Topic 4)}
$. ***$p$<0.001, **$p$<0.01, *$p$<0.05, $\dag p$<0.10.}
\label{table:regression-confidence}
\end{table}

\begin{table}[!ht]
\centering
\footnotesize
\begin{tabular}[t]{llSSSccc}
\toprule
\textbf{Model} & \textbf{Variable} & \textbf{b} & \textbf{S.E.} & \textbf{$p$} & \textbf{95\% CI Lower} & \textbf{95\% CI Upper}\\
\midrule
\multirow{4}{*}{Model 1} & (Intercept) & 41.67*** & 1.96 & <0.001 & 37.82 & 45.53\\
 & AI Label & -1.66 & 1.96 & 0.396 & -5.49 & 2.18\\
 & No Label & -0.00 & 1.96 & 0.998 & -3.84 & 3.83\\
 & $R^2$=0.011 & & & & \\
 \midrule
\multirow{4}{*}{Model 2} & (Intercept) & 41.67*** & 1.96 & <0.001 & 37.82 & 45.52\\
 & AI Label & -1.65 & 1.96 & 0.398 & -5.49 & 2.19\\
 & Human Label & 0.00 & 1.96 & 0.998 & -3.83 & 3.84\\
 & $R^2$=0.011 & & & & \\
\bottomrule
\end{tabular}
\caption{Regression coefficients for the sharing intention variable. Model 1 treats the \textit{human label} condition as the reference category, given by the regression equation $
\text{Post-Intervention Sharing} = \beta_0 + \beta_1 \text{(AI Label Condition)} + \beta_2 \text{(No Label Condition)} + \beta_3 \text{(Topic 2)} + \beta_4 \text{(Topic 3)} + \beta_5 \text{(Topic 4)}
$.
Model 2 treats the \textit{no label} condition as the reference category, given by the regression equation
$
\text{Post-Intervention Sharing} = \beta_0 + \beta_1 \text{(AI Label Condition)} + \beta_2 \text{(Human Label Condition)} + \beta_3 \text{(Topic 2)} + \beta_4 \text{(Topic 3)} + \beta_5 \text{(Topic 4)}
$. ***$p$<0.001, **$p$<0.01, *$p$<0.05, $\dag p$<0.10.}
\label{table:regression-sharing}
\end{table}

\begin{table}[!ht]
\centering
\footnotesize
\begin{tabular}[t]{llSSSccc}
\toprule
\textbf{Model} & \textbf{Variable} & \textbf{b} & \textbf{S.E.} & \textbf{$p$} & \textbf{95\% CI Lower} & \textbf{95\% CI Upper}\\
\midrule
\multirow{4}{*}{Model 1} & (Intercept) & 60.82*** & 1.18 & <0.001 & 58.50 & 63.15\\
 & AI Label & -1.19 & 1.18 & 0.315 & -3.50 & 1.13\\
 & No Label & -0.83 & 1.18 & 0.479 & -3.15 & 1.48\\
 & $R^2$=0.020 & & & & \\
 \midrule
\multirow{4}{*}{Model 2} & (Intercept) & 59.99*** & 1.18 & <0.001 & 57.67 & 62.31\\
 & AI Label & -0.35 & 1.18 & 0.766 & -2.67 & 1.96\\
 & Human Label & 0.83 & 1.18 & 0.479 & -1.48 & 3.15\\
 & $R^2$=0.020 & & & & \\
\bottomrule
\end{tabular}
\caption{Regression coefficients for the accuracy judgment variable. Model 1 treats the \textit{human label} condition as the reference category, given by the regression equation $
\text{Post-Intervention Accuracy} = \beta_0 + \beta_1 \text{(AI Label Condition)} + \beta_2 \text{(No Label Condition)} + \beta_3 \text{(Topic 2)} + \beta_4 \text{(Topic 3)} + \beta_5 \text{(Topic 4)}
$.
Model 2 treats the \textit{no label} condition as the reference category, given by the regression equation
$
\text{Post-Intervention Accuracy} = \beta_0 + \beta_1 \text{(AI Label Condition)} + \beta_2 \text{(Human Label Condition)} + \beta_3 \text{(Topic 2)} + \beta_4 \text{(Topic 3)} + \beta_5 \text{(Topic 4)}
$. ***$p$<0.001, **$p$<0.01, *$p$<0.05, $\dag p$<0.10.}
\label{table:regression-accuracy}
\end{table}

\begin{table}[!ht]
\centering
\scriptsize
\begin{tabular}{llSScccc}
\toprule
\textbf{Model} & \textbf{Variable} & \textbf{b} & \textbf{S.E.} & \textbf{$p$ (BH Adjusted)} & \textbf{95\% CI Lower} & \textbf{95\% CI Upper}\\
\midrule
\multirow{11}{*}{Model 1: Political Party} & (Intercept) & 22.95*** & 3.27 & <0.001 & 16.53 & 29.36\\
 & Pre-Intervention DV & 0.83*** & 0.01 & <0.001 & 0.80 & 0.86\\
 & AI Label & -1.44 & 4.07 & 0.723 & -9.42 & 6.54\\
 & No Label & 1.44 & 3.96 & 0.716 & -6.32 & 9.21\\
 & Democrat x AI Label & -1.35 & 4.29 & 0.754 & -9.77 & 7.07\\
 & Republican x AI Label & 4.12 & 4.55 & 0.366 & -4.81 & 13.05\\
 & Independent x AI Label & 0.57 & 4.47 & 0.898 & -8.19 & 9.34\\
 & Democrat x No Label & -3.14 & 4.18 & 0.452 & -11.34 & 5.05\\
 & Republican x No Label & -3.47 & 4.50 & 0.441 & -12.30 & 5.36\\
 & Independent x No Label & -2.13 & 4.39 & 0.627 & -10.74 & 6.47\\
 & $R^2$=0.679 & & & & & \\
\midrule
 & (Intercept) & 26.78*** & 1.68 & <0.001 & 23.48 & 30.08\\
Model 1: Democrat & Pre-Intervention DV & 0.81*** & 0.02 & <0.001 & 0.77 & 0.85\\
\textit{Subgroup} & AI Label & -2.85 & 1.37 & 0.038 (0.114) & -5.53 & -0.16\\
& No Label & -1.68 & 1.34 & 0.210 & -4.32 & 0.95\\
 & $R^2$=0.649 & & & & & \\
\midrule
 & (Intercept) & 23.48*** & 3.02 & <0.001 & 17.54 & 29.43\\
Model 1: Republican & Pre-Intervention DV & 0.80*** & 0.03 & <0.001 & 0.74 & 0.87\\
\textit{Subgroup} & AI Label & 2.52 & 2.26 & 0.264 (0.396) & -1.92 & 6.96\\
 & No Label & -2.17 & 2.34 & 0.354 & -6.79 & 2.44\\
 & $R^2$=0.681 & & & & & \\
\midrule
 & (Intercept) & 22.71*** & 2.15 & <0.001 & 18.48 & 26.94\\
Model 1: Independent & Pre-Intervention DV & 0.90*** & 0.03 & <0.001 & 0.84 & 0.95\\
\textit{Subgroup} & AI Label & -0.98 & 1.67 & 0.560 (0.560) & -4.26 & 2.31\\
 & No Label & -0.54 & 1.70 & 0.750 & -3.88 & 2.80\\
 & $R^2$=0.749 & & & & & \\
\midrule
\multirow{11}{*}{Model 2: Political Party}  & (Intercept) & 24.39*** & 2.61 & <0.001 & 19.28 & 29.50\\
 & Pre-Intervention DV & 0.83*** & 0.01 & <0.001 & 0.80 & 0.86\\
 & AI Label & -2.88 & 3.55 & 0.417 & -9.85 & 4.08\\
 & Human Label & -1.44 & 3.96 & 0.716 & -9.21 & 6.32\\
 & Democrat x AI Label & 1.80 & 3.79 & 0.636 & -5.64 & 9.23\\
 & Republican x AI Label & 7.59$\dag$ & 4.13 & 0.066 & -0.51 & 15.70\\
 & Independent x AI Label & 2.71 & 4.06 & 0.506 & -5.27 & 10.68\\
 & Democrat x Human Label & 3.14 & 4.18 & 0.452 & -5.05 & 11.34\\
 & Republican x Human Label & 3.47 & 4.50 & 0.441 & -5.36 & 12.30\\
 & Independent x Human Label & 2.13 & 4.39 & 0.627 & -6.47 & 10.74\\
 & $R^2$=0.679 & & & & & \\
\midrule
 & (Intercept) & 25.10*** & 1.70 & <0.001 & 21.77 & 28.42\\
Model 2: Democrat & Pre-Intervention DV & 0.81*** & 0.02 & <0.001 & 0.77 & 0.85\\
\textit{Subgroup} & AI Label & -1.16 & 1.33 & 0.381 (0.572) & -3.77 & 1.44\\
 & Human Label & 1.68 & 1.34 & 0.210 & -0.95 & 4.32\\
 & $R^2$=0.649 & & & & & \\
\midrule
 & (Intercept) & 21.31*** & 3.12 & <0.001 & 15.17 & 27.44\\
Model 2: Republican & Pre-Intervention DV & 0.80*** & 0.03 & <0.001 & 0.74 & 0.87\\
\textit{Subgroup} & AI Label & 4.70 & 2.31 & 0.043 (0.129) & 0.15 & 9.25\\
 & Human Label & 2.17 & 2.34 & 0.354 & -2.44 & 6.79\\
 & $R^2$=0.681 & & & & & \\
\midrule
 & (Intercept) & 22.17*** & 2.19 & <0.001 & 17.86 & 26.48\\
Model 2: Independent & Pre-Intervention DV & 0.90*** & 0.03 & <0.001 & 0.84 & 0.95\\
\textit{Subgroup} & AI Label & -0.43 & 1.79 & 0.808 (0.808) & -3.95 & 3.08\\
 & Human Label & 0.54 & 1.70 & 0.750 & -2.80 & 3.88\\
 & $R^2$=0.749 & & & & & \\
\bottomrule
\end{tabular}
\caption{Regression coefficients for the support variable, with interaction terms for political party identity.
An interaction term is added between each of the labeling condition dummy variables. Subgroup analyses do not include any interaction terms, but perform the regression on a subset of the population. Benjamini–Hochberg adjusted $p$-values for the \textit{AI label} condition are reported in parentheses. ***$p$<0.001, **$p$<0.01, *$p$<0.05, $\dag p$<0.10.}
\label{table:regression-party}
\end{table}

\begin{table}[!ht]
\centering
\scriptsize
\begin{tabular}{llSSccc}
\toprule
\textbf{Model} & \textbf{Variable} & \textbf{b} & \textbf{S.E.} & \textbf{$p$ (BH Adjusted)} & \textbf{95\% CI Lower} & \textbf{95\% CI Upper}\\
\midrule
\multirow{7}{*}{Model 1: Prior Knowledge}  & (Intercept) & 24.72*** & 1.31 & <0.001 & 22.15 & 27.30\\
 & Pre-Intervention DV & 0.84*** & 0.01 & <0.001 & 0.81 & 0.87\\
 & AI Label & 0.17 & 1.30 & 0.893 & -2.37 & 2.72\\
 & No Label & -0.90 & 1.31 & 0.492 & -3.47 & 1.67\\
 & Knowledge x AI Label & -5.62 & 4.00 & 0.161 & -13.47 & 2.24\\
 & Knowledge x No Label & -1.24 & 4.08 & 0.761 & -9.24 & 6.76\\
 & $R^2$=0.680 & & & & & \\
\midrule
 & (Intercept) & 25.69*** & 1.34 & <0.001 & 23.05 & 28.32\\
Model 1: Low Knowledge & Pre-Intervention DV & 0.83*** & 0.02 & <0.001 & 0.79 & 0.86\\
\textit{Subgroup} & AI Label & -0.48 & 1.08 & 0.660 (0.660) & -2.59 & 1.64\\
 & No Label & -1.01 & 1.07 & 0.349 & -3.12 & 1.10\\
 & $R^2$=0.649 & & & & & \\
\midrule
 & (Intercept) & 13.93*** & 3.38 & <0.001 & 7.27 & 20.59\\
Model 1: Moderate Knowledge & Pre-Intervention DV & 0.89*** & 0.03 & <0.001 & 0.83 & 0.95\\
\textit{Subgroup} & AI Label & -2.67 & 2.06 & 0.195 (0.585) & -6.72 & 1.38\\
 & No Label & -1.71 & 2.11 & 0.418 & -5.87 & 2.44\\
 & $R^2$=0.778 & & & & & \\
\midrule
 & (Intercept) & 20.02** & 7.53 & 0.009 & 5.06 & 34.97\\
Model 1: High Knowledge & Pre-Intervention DV & 0.82*** & 0.06 & <0.001 & 0.71 & 0.94\\
\textit{Subgroup} & AI Label & -3.16 & 4.40 & 0.474 (0.660) & -11.90 & 5.58\\
 & No Label & -0.87 & 4.38 & 0.843 & -9.57 & 7.83\\
 & $R^2$=0.703 & & & & & \\
\midrule
\multirow{7}{*}{Model 2: Prior Knowledge} & (Intercept) & 23.82*** & 1.33 & <0.001 & 21.21 & 26.44\\
 & Pre-Intervention DV & 0.84*** & 0.01 & <0.001 & 0.81 & 0.87\\
 & AI Label & 1.07 & 1.30 & 0.408 & -1.47 & 3.62\\
 & Human Label & 0.90 & 1.31 & 0.492 & -1.67 & 3.47\\
 & Knowledge x AI Label & -4.38 & 4.05 & 0.280 & -12.31 & 3.56\\
 & Knowledge x Human Label & 1.24 & 4.08 & 0.761 & -6.76 & 9.24\\
 & $R^2$=0.680 & & & & & \\
\midrule
 & (Intercept) & 24.68*** & 1.36 & <0.001 & 22.01 & 27.35\\
Model 2: Low Knowledge & Pre-Intervention DV & 0.83*** & 0.02 & <0.001 & 0.79 & 0.86\\
\textit{Subgroup} & AI Label & 0.53 & 1.08 & 0.620 (0.649) & -1.58 & 2.64\\
 & Human Label & 1.01 & 1.07 & 0.349 & -1.10 & 3.12\\
 & $R^2$=0.649 & & & & & \\
\midrule
 & (Intercept) & 12.22*** & 3.42 & <0.001 & 5.48 & 18.95\\
Model 2: Moderate Knowledge & Pre-Intervention DV & 0.89*** & 0.03 & <0.001 & 0.83 & 0.95\\
\textit{Subgroup} & AI Label & -0.96 & 2.11 & 0.649 (0.649) & -5.11 & 3.19\\
 & Human Label & 1.71 & 2.11 & 0.418 & -2.44 & 5.87\\
 & $R^2$=0.778 & & & & & \\
\midrule
 & (Intercept) & 19.15* & 7.69 & 0.015 & 3.86 & 34.43\\
Model 2: High Knowledge & Pre-Intervention DV & 0.82*** & 0.06 & <0.001 & 0.71 & 0.94\\
\textit{Subgroup} & AI Label & -2.29 & 4.34 & 0.599 (0.649) & -10.93 & 6.34\\
 & Human Label & 0.87 & 4.38 & 0.843 & -7.83 & 9.57\\
 & $R^2$=0.703 & & & & & \\
\bottomrule
\end{tabular}
\caption{Regression coefficients for the support variable, with interaction terms for prior knowledge about the topic.
An interaction term is added between each of the labeling condition dummy variables. Subgroup analyses do not include any interaction terms, but perform the regression on a subset of the population. Benjamini–Hochberg adjusted $p$-values for the \textit{AI label} condition are reported in parentheses. ***$p$<0.001, **$p$<0.01, *$p$<0.05, $\dag p$<0.10.}
\label{table:regression-knowledge}
\end{table}

\begin{table}[!ht]
\centering
\scriptsize
\begin{tabular}{llSScccc}
\toprule
\textbf{Model} & \textbf{Variable} & \textbf{b} & \textbf{S.E.} & \textbf{$p$ (BH Adjusted)} & \textbf{95\% CI Lower} & \textbf{95\% CI Upper}\\
\midrule
\multirow{7}{*}{Model 1: AI Experience} & (Intercept) & 25.14*** & 1.76 & <0.001 & 21.69 & 28.59\\
 & Pre-Intervention DV & 0.83*** & 0.01 & <0.001 & 0.80 & 0.86\\
 & AI Label & -1.23 & 2.15 & 0.567 & -5.44 & 2.98\\
 & No Label & -1.12 & 2.12 & 0.599 & -5.28 & 3.05\\
 & Experience x AI Label & 0.33 & 3.29 & 0.920 & -6.12 & 6.78\\
 & Experience x No Label & -0.07 & 3.28 & 0.984 & -6.51 & 6.38\\
 & $R^2$=0.677 & & & & \\
\midrule
 & (Intercept) & 26.26*** & 1.56 & <0.001 & 23.20 & 29.33\\
Model 1: Low Experience & Pre-Intervention DV & 0.83*** & 0.02 & <0.001 & 0.79 & 0.87\\
\textit{Subgroup} & AI Label & -1.61 & 1.29 & 0.214 (0.428) & -4.15 & 0.93\\
 & No Label & -1.40 & 1.27 & 0.270 & -3.90 & 1.10\\
 & $R^2$=0.716 & & & & \\
\midrule
 & (Intercept)35 & 23.28*** & 1.78 & <0.001 & 19.78 & 26.78\\
Model 1: High Experience & Pre-Intervention DV & 0.83*** & 0.02 & <0.001 & 0.79 & 0.88\\
\textit{Subgroup} & AI Label & -0.40 & 1.35 & 0.765 (0.765) & -3.06 & 2.25\\
 & No Label & -0.84 & 1.37 & 0.537 & -3.52 & 1.84\\
 & $R^2$=0.647 & & & & \\
\midrule
\multirow{7}{*}{Model 2: AI Experience} & (Intercept) & 24.02*** & 1.82 & <0.001 & 20.44 & 27.60\\
 & Pre-Intervention DV & 0.83*** & 0.01 & <0.001 & 0.80 & 0.86\\
 & AI Label & -0.11 & 2.17 & 0.958 & -4.36 & 4.14\\
 & Human Label & 1.12 & 2.12 & 0.599 & -3.05 & 5.28\\
 & Experience x AI Label & 0.40 & 3.31 & 0.905 & -6.10 & 6.90\\
 & Experience x Human Label & 0.07 & 3.28 & 0.984 & -6.38 & 6.51\\
 & $R^2$=0.677 & & & & \\
\midrule
 & (Intercept) & 24.86*** & 1.63 & <0.001 & 21.65 & 28.06\\
Model 2: Low Experience & Pre-Intervention DV & 0.83*** & 0.02 & <0.001 & 0.79 & 0.87\\
\textit{Subgroup} & AI Label & -0.20 & 1.30 & 0.875 (0.875) & -2.76 & 2.35\\
 & Human Label & 1.40 & 1.27 & 0.270 & -1.10 & 3.90\\
 & $R^2$=0.716 & & & & \\
\midrule
 & (Intercept) & 22.44*** & 1.75 & <0.001 & 19.00 & 25.87\\
Model 2: High Experience & Pre-Intervention DV & 0.83*** & 0.02 & <0.001 & 0.79 & 0.88\\
\textit{Subgroup} & AI Label & 0.44 & 1.35 & 0.745 (0.875) & -2.21 & 3.08\\
 & Human Label & 0.84 & 1.37 & 0.537 & -1.84 & 3.52\\
 & $R^2$=0.647 & & & & \\
\bottomrule
\end{tabular}
\caption{Regression coefficients for the support variable, with interaction terms for prior experience with AI tools.
An interaction term is added between each of the labeling condition dummy variables. Subgroup analyses do not include any interaction terms, but perform the regression on a subset of the population. Benjamini–Hochberg adjusted $p$-values for the \textit{AI label} condition are reported in parentheses. ***$p$<0.001, **$p$<0.01, *$p$<0.05, $\dag p$<0.10.}
\label{table:regression-experience}
\end{table}

\begin{table}[!ht]
\centering
\scriptsize
\begin{tabular}{llSScccc}
\toprule
\textbf{Model} & \textbf{Variable} & \textbf{b} & \textbf{S.E.} & \textbf{$p$ (BH Adjusted)} & \textbf{95\% CI Lower} & \textbf{95\% CI Upper}\\
\midrule
\multirow{7}{*}{Model 1: Education} & (Intercept) & 24.40*** & 2.11 & <0.001 & 20.27 & 28.53\\
 & Pre-Intervention DV & 0.83*** & 0.01 & <0.001 & 0.80 & 0.86\\
 & AI Label & 3.35 & 2.68 & 0.211 & -1.91 & 8.61\\
 & No Label & -1.45 & 2.74 & 0.597 & -6.82 & 3.92\\
 & Education x AI Label & -7.38$\dag$ & 4.19 & 0.079 & -15.60 & 0.85\\
 & Education x No Label & 0.54 & 4.19 & 0.897 & -7.67 & 8.76\\
 & $R^2$=0.679 & & & & \\
\midrule
 & (Intercept) & 25.25*** & 1.72 & <0.001 & 21.88 & 28.63\\
Model 1: Less than Bachelor's & Pre-Intervention DV & 0.82*** & 0.02 & <0.001 & 0.78 & 0.86\\
\textit{Subgroup} & AI Label & 0.56 & 1.37 & 0.681 (0.681) & -2.13 & 3.25\\
 & No Label & -0.40 & 1.42 & 0.780 & -3.18 & 2.39\\
 & $R^2$=0.698 & & & & \\
\midrule
 & (Intercept) & 24.33*** & 1.64 & <0.001 & 21.11 & 27.55\\
Model 1: Bachelor's or more & Pre-Intervention DV & 0.84*** & 0.02 & <0.001 & 0.81 & 0.88\\
\textit{Subgroup} & AI Label & -2.00 & 1.28 & 0.120 (0.240) & -4.52 & 0.52\\
 & No Label & -1.29 & 1.26 & 0.305 & -3.75 & 1.18\\
 & $R^2$=0.667 & & & & \\
\midrule
\multirow{7}{*}{Model 2: Education} & (Intercept) & 22.95*** & 2.20 & <0.001 & 18.64 & 27.26\\
 & Pre-Intervention DV & 0.83*** & 0.01 & <0.001 & 0.80 & 0.86\\
 & AI Label & 4.80$\dag$ & 2.74 & 0.080 & -0.58 & 10.18\\
 & Human Label & 1.45 & 2.74 & 0.597 & -3.92 & 6.82\\
 & Education x AI Label & -7.92$\dag$ & 4.23 & 0.061 & -16.22 & 0.38\\
 & Education x Human Label & -0.54 & 4.19 & 0.897 & -8.76 & 7.67\\
 & $R^2$=0.679 & & & & \\
\midrule
 & (Intercept) & 24.86*** & 1.79 & <0.001 & 21.34 & 28.37\\
Model 2: Less than Bachelor's & Pre-Intervention DV & 0.82*** & 0.02 & <0.001 & 0.78 & 0.86\\
\textit{Subgroup} & AI Label & 0.96 & 1.40 & 0.494 (0.576) & -1.79 & 3.71\\
 & Human Label & 0.40 & 1.42 & 0.780 & -2.39 & 3.18\\
 & $R^2$=0.698 & & & & \\
\midrule
 & (Intercept) & 23.04*** & 1.63 & <0.001 & 19.85 & 26.23\\
Model 2: Bachelor's or more & Pre-Intervention DV & 0.84*** & 0.02 & <0.001 & 0.81 & 0.88\\
\textit{Subgroup} & AI Label & -0.71 & 1.27 & 0.576 (0.576) & -3.20 & 1.78\\
 & Human Label & 1.29 & 1.26 & 0.305 & -1.18 & 3.75\\
 & $R^2$=0.667 & & & & \\
\bottomrule
\end{tabular}
\caption{Regression coefficients for the support variable, with interaction terms for education level.
An interaction term is added between each of the labeling condition dummy variables. Subgroup analyses do not include any interaction terms, but perform the regression on a subset of the population. Benjamini–Hochberg adjusted $p$-values for the \textit{AI label} condition are reported in parentheses. ***$p$<0.001, **$p$<0.01, *$p$<0.05, $\dag p$<0.10.}
\label{table:regression-education}
\end{table}

\begin{table}[!ht]
\centering
\scriptsize
\begin{tabular}{llSSccc}
\toprule
\textbf{Model} & \textbf{Variable} & \textbf{b} & \textbf{S.E.} & \textbf{$p$ (BH Adjusted)} & \textbf{95\% CI Lower} & \textbf{95\% CI Upper}\\
\midrule
\multirow{7}{*}{Model: Age} & (Intercept) & 24.93*** & 2.46 & <0.001 & 20.09 & 29.76\\
 & Pre-Intervention DV & 0.83*** & 0.01 & <0.001 & 0.80 & 0.86\\
 & AI Label & 4.44 & 3.18 & 0.163 & -1.80 & 10.69\\
 & No Label & 0.42 & 3.22 & 0.897 & -5.90 & 6.73\\
 & Age x AI Label & -0.14$\dag$ & 0.08 & 0.073 & -0.29 & 0.01\\
 & Age x No Label & -0.04 & 0.08 & 0.612 & -0.19 & 0.11\\
 & $R^2$=0.678 & & & & & \\
\midrule
 & (Intercept) & 24.15*** & 1.65 & <0.001 & 20.92 & 27.39\\
Model 1: Below Median Age & Pre-Intervention DV & 0.83*** & 0.02 & <0.001 & 0.79 & 0.87\\
\textit{Subgroup} & AI Label & 1.14 & 1.29 & 0.380 (0.380) & -1.41 & 3.68\\
 & No Label & -0.56 & 1.29 & 0.662 & -3.10 & 1.97\\
 & $R^2$=0.659 & & & & & \\
\midrule
 & (Intercept) & 25.52*** & 1.71 & <0.001 & 22.15 & 28.88\\
Model 1: Above Median Age & Pre-Intervention DV & 0.83*** & 0.02 & <0.001 & 0.79 & 0.87\\
\textit{Subgroup} & AI Label & -3.13* & 1.36 & 0.022 (0.044) & -5.80 & -0.46\\
 & No Label & -1.77 & 1.36 & 0.196 & -4.45 & 0.91\\
 & $R^2$=0.693 & & & & & \\
\midrule
\multirow{7}{*}{Model 2: Age} & (Intercept) & 25.34*** & 2.51 & <0.001 & 20.42 & 30.27\\
 & Pre-Intervention DV & 0.83*** & 0.01 & <0.001 & 0.80 & 0.86\\
 & AI Label & 4.02 & 3.22 & 0.211 & -2.29 & 10.34\\
 & Human Label & -0.42 & 3.22 & 0.897 & -6.73 & 5.90\\
 & Age x AI Label & -0.10 & 0.08 & 0.204 & -0.25 & 0.05\\
 & Age x Human Label & 0.04 & 0.08 & 0.612 & -0.11 & 0.19\\
 & $R^2$=0.678 & & & & & \\
\midrule
 & (Intercept) & 23.59*** & 1.67 & <0.001 & 20.31 & 26.87\\
Model 2: Below Median Age & Pre-Intervention DV & 0.83*** & 0.02 & <0.001 & 0.79 & 0.87\\
\textit{Subgroup} & AI Label & 1.70 & 1.31 & 0.194 (0.314) & -0.87 & 4.27\\
 & Human Label & 0.56 & 1.29 & 0.662 & -1.97 & 3.10\\
 & $R^2$=0.660 & & & & & \\
\midrule
 & (Intercept) & 23.75*** & 1.73 & <0.001 & 20.35 & 27.15\\
Model 2: Above Median Age & Pre-Intervention DV & 0.83*** & 0.02 & <0.001 & 0.79 & 0.87\\
\textit{Subgroup} & AI Label & -1.36 & 1.35 & 0.314 (0.314) & -4.01 & 1.29\\
 & Human Label & 1.77 & 1.36 & 0.196 & -0.91 & 4.45\\
 & $R^2$=0.693 & & & & & \\
\bottomrule
\end{tabular}
\caption{Regression coefficients for the support variable, with interaction terms for age.
An interaction term is added between each of the labeling condition dummy variables. Subgroup analyses do not include any interaction terms, but perform the regression on a subset of the population. Benjamini–Hochberg adjusted $p$-values for the \textit{AI label} condition are reported in parentheses. ***$p$<0.001, **$p$<0.01, *$p$<0.05, $\dag p$<0.10.}
\label{table:regression-age}
\end{table}

\end{document}